\documentclass[twocolumn,english,prl,showpacs]{revtex4}
\usepackage[colorlinks=true,urlcolor=blue,citecolor=blue,linkcolor=blue]{hyperref}
\usepackage[T1]{fontenc}
\usepackage[latin9]{inputenc}
\usepackage{amssymb}
\usepackage{graphicx}
\usepackage{amsmath,color}
\usepackage{mathrsfs}
\usepackage{float}
\usepackage{indentfirst}
\usepackage{amsmath}
\usepackage{subfigure}
\DeclareMathOperator{\Tr}{Tr}

\makeatletter

\usepackage{tikz}
\usetikzlibrary{arrows}
\usetikzlibrary{calc}
\usepackage{babel}

\begin{document}

\title{Renyi Entanglement Entropy of Interacting Fermions Calculated Using \\Continuous-Time Quantum Monte Carlo Method}

\author{Lei Wang and Matthias Troyer}
\affiliation{Theoretische Physik, ETH Zurich, 8093 Zurich, Switzerland}

\begin{abstract}
We present a new algorithm for calculating the Renyi entanglement entropy of interacting fermions using the continuous-time quantum Monte Carlo method. The algorithm only samples  \emph{interaction correction} of the entanglement entropy, which by design ensures efficient calculation of  weakly interacting systems. Combined with Monte Carlo reweighting, the algorithm also performs well for systems with strong interactions. We demonstrate the potential of this method by studying the quantum entanglement signatures of the charge-density-wave transition of interacting fermions on a square lattice.  
\end{abstract}


\maketitle

We have entered an information age of condensed matter physics. Many information theoretical tools have been used to identify exotic phases and phase transitions~\cite{Feiguin:2007bo, PhysRevLett.99.140405, Isakov:2011fz, Jiang:2012dw, PhysRevLett.109.067201}. These tools are especially useful when a conventional local order parameter characterization fails, e.g. in the case of topological order~\cite{Wen:1990tm}. The reduced density matrix $\hat{\rho}_{A}$ plays a central role in this quantum information perspective. It contains all the information about a subregion A when viewing the remaining part of the system as environment. The entanglement entropy (EE) quantitatively measures the entanglement between the subregion A and its environment. The rank-$n$ Renyi EE, in particular, reads

\begin{equation}
S_{n} = \frac{1}{1-n}\ln \left[\Tr (\hat{\rho}_{A}^{n})\right].
\label{eq:REE}
\end{equation}
Compared to the more familiar von Neumann entropy the Renyi EE is easier to evaluate both analytically and numerically. Quantum Monte Carlo methods have been used to calculate entanglement properties of many bosonic~\cite{Buividovich:2008hz, Hastings:2010dca, Herdman:2014jqa} and fermionic~\cite{Zhang:2011ka, McMinis:2013dp, Grover:2013cs, Broecker:2014ud} systems. Some interesting applications include the calculation of the topological entanglement entropy~\cite{Levin:2006ij, Kitaev:2006dn, Isakov:2011fz} and the identification of interacting topological insulators~\cite{2014PhRvB..89l5121A, Anonymous:XiXakTeu}. 

Since the Monte Carlo method samples $\Tr (\hat{\rho}_{A}^{n})$ but not $S_{n}$ directly, it will suffer from a severe problem when the entanglement entropy is large~\cite{Hastings:2010dca, Zhang:2011ka, Grover:2013cs}. For example, the Renyi EE of a generic gapped many-body system follows an area law~\cite{Eisert:2010hq} and as $S_{n}$ increases linearly with the boundary size, the Monte Carlo simulation has to sample rare events whose probability vanishes exponentially. The problem is even more severe for \emph{highly entangled} systems which violate the area law~\cite{PhysRevLett.96.010404, PhysRevLett.96.100503, PhysRevLett.105.050502,PhysRevX.2.011012, PhysRevLett.111.210402,PhysRevLett.112.160403}, including the systems with Fermi surface or at nonzero temperature. One way to alleviate the problem is to employ the ratio method~\cite{Hastings:2010dca, Humeniuk:2012cq}, which splits the estimator of $\Tr(\hat{\rho}_{A}^{n})$ into products of intermediate ratios and samples each term individually. However, too many intermediate ratios will lead to an accumulation of errors in the final result~\cite{Humeniuk:2012cq}. 

In this paper we introduce an algorithm for computing the Renyi EE of interacting fermions using the continuous-time quantum Monte Carlo method (CTQMC)~\cite{Rubtsov:2005iw, Gull:2011jd}. The main advantage of the algorithm is that it samples the \emph{interaction correction} of the Renyi EE. By design the method allows to easily calculate large Renyi EE of  weakly interacting systems. Our benchmark shows that it is also applicable to strongly interacting systems when combining  with the ratio method~\cite{Hastings:2010dca, Humeniuk:2012cq}. As an application we calculate the Renyi EE of spinless fermions across a charge-density-wave transition and show it correctly predicts the critical point. 

In the replica approach~\cite{Calabrese:2004hl, Melko:2010jda}, the rank-$n$ Renyi entanglement entropy is calculated as 

\begin{equation}
S_{n} = \frac{1}{1-n}\ln\left(\frac{\mathcal{Z}^{A}}{\mathcal{Z}^{n}}\right)
\label{eq:S2}
\end{equation}
where $\mathcal{Z}= \Tr(e^{-\beta \hat{H}})$ is the partition 
function at inverse temperature $\beta=1/T$ and $\mathcal{Z}^{A}$ is the partition function defined on a $n$-sheeted Riemann surface~\cite{Melko:2010jda}. Ref.\cite{Broecker:2014ud} introduced an artificial system with imaginary-time dependent Hamiltonian whose partition function at inverse temperature $n\beta$ equals to $\mathcal{Z}^{A}$. We here formulate an efficient algorithm to calculate Eq.~({\ref{eq:S2}}) using CTQMC. 



The first and crucial step of our method is to split Eq.~(\ref{eq:S2}) into  $S_{n} = S_{n}^{0} + \Delta S_{n}$, in which 

\begin{eqnarray}
S_{n}^{0}  & =& \frac{1}{1-n} \ln\left(\frac{\mathcal{Z}_{0}^{A}}{\mathcal{Z}_{0}^{n}}\right), \\ 
\Delta S_{n} & = & \frac{1}{1-n} \left\{ \ln\left[\eta \frac{\mathcal{Z}^{A}/\mathcal{Z}_{0}^{A}}{ (\mathcal{Z}/ \mathcal{Z}_{0}  )^{n}}\right] - \ln(\eta) \right\}.    
\label{eq:dS2}
\end{eqnarray}
$\mathcal{Z}_{0}$ and $\mathcal{Z}_{0}^{A}$ are the noninteracting counterparts of $\mathcal{Z}$ and $\mathcal{Z}^{A}$~\footnote{In general one can choose $\mathcal{Z}_{0}$ and $\mathcal{Z}^{A}_{0}$ as partition functions of any quadratic Hamiltonian. It amounts to redefine the reference Hamiltonian around which one performs the interaction expansion. A particular useful example is to expand around a mean-field Hamiltonian and the resulting $\Delta S_{n}$ will be the entanglement entropy difference to a mean-field state. Employing this trick will further improve efficiency of our algorithm, especially in the long-range-ordered phase. We thank Fakher Assaad for pointing this out to us.}. $S_{n}^{0} $ is the Renyi EE of noninteracting fermions and can be calculated easily using the correlation matrix method~\cite{Peschel:2002gz, SM}. Eq.~(\ref{eq:dS2}) describes the \emph{interaction corrections} to the Renyi EE which is presumably much smaller than $S_{n}^{0}$ for weakly interacting systems. We use Monte Carlo (MC) sampling to calculate the quantity in the square bracket of Eq.~(\ref{eq:dS2}), where we have introduced a free parameter $\eta$ to further control the MC dynamics. For an optimal choice of $\eta$ the sampled quantity in the square bracket is close to one and  $\frac{-1}{1-n}\ln(\eta)$ will contribute mostly to $\Delta S_{n}$. The fact that MC sampling corrects an educated guess is another appealing feature of the present algorithm. In practice, $\eta$ can be determined from a rough estimate of $\Delta S_{n}$ (either based on existing theory of EE scaling laws or in the MC equilibration steps). MC sampling will correct the estimate and restore the exact result no matter what the initial choice of $\eta$ was. 

\begin{figure}[t]
  \centering
\subfigure[]
{
\begin{tikzpicture}[]
  \coordinate (i) at (0,0);
  \coordinate (j) at (1.6,0);   
  \coordinate (k) at (3.2,0); 
  \draw[|-|] (i) node [left] {$0$} -- (j) node [below] {$\beta$};
  \draw[-|] (j) node [left] {} -- (k) node [right] {$2\beta$};
 
  \coordinate (i) at (0.3,0.8);
  \coordinate (j) at (0.3,-0.8);
  \draw [-](i) -- (j) node [below] {};
  \shade[ball color=red] (i) circle (1mm) node [right] {$\mathbf{i}_{1}$};
  \shade[ball color=blue] (j) circle (1mm) node [right] {$\mathbf{i}_{2}$};
  \draw (0.3 ,0) circle (0.5mm) node [below right] {$\tau_{2}$};
   \draw (0.3 ,0) node [above right] {$\tau_{1}$};
    
  \coordinate (i) at (0.9,0.8);
  \coordinate (j) at (0.9,-0.8);
  \draw [-](i) -- (j) node [below] {};
  \shade[ball color=blue] (i) circle (1mm) node [right] {$\mathbf{i}_{3}$};
  \shade[ball color=red] (j) circle (1mm) node [right] {$\mathbf{i}_{4}$};
  \draw (0.9,0) circle (0.5mm) node [below right] {$\tau_{4}$};
    \draw (0.9,0)  node [above right] {$\tau_{3}$};

  \coordinate (i) at (2.4,0.8);
  \coordinate (j) at (2.4,-0.8);
  \draw [-](i) -- (j) node [below] {};
  \shade[ball color=red] (i) circle (1mm) node [right] {$\mathbf{i}_{5}$};
  \shade[ball color=blue] (j) circle (1mm) node [right] {$\mathbf{i}_{6}$};
  \draw (2.4 ,0) circle (0.5mm) node [below right] {$\tau_{6}$};
    \draw (2.4 ,0) circle (0.5mm) node [above right] {$\tau_{5}$};
\end{tikzpicture}
} 
\subfigure[]
{
\begin{tikzpicture}[->,>=stealth',shorten >=1pt,auto,
  thick,main node/.style={circle,fill=blue!20,draw,font=\sffamily\large\bfseries}]  

  \node[main node]         (Z2) at (-1.1,0) {$\mathcal{Z}^{2}$};
  \node[main node]         (ZA) at (1.1,0) {$\mathcal{Z}^{A}$};

 \path[every node/.style={font=\sffamily\scriptsize}]
      (Z2)   edge [bend right]  node[below] {Eq.~(\ref{eq:switch})} (ZA)
             edge [loop above]  node {Eqs.~(\ref{eq:add}-\ref{eq:remove})} (Z2)
              
      (ZA)   edge [bend right]  node[above] {Eq.~(\ref{eq:switch})} (Z2)
             edge [loop above]  node {Eqs.~(\ref{eq:add}-\ref{eq:remove})} (ZA); 
\end{tikzpicture}
} 

\caption{{Key concepts of the algorithm.} (a) A configuration with $k_{1}=2$ vertices in the time interval $[0, \beta)$ and $k_{2}=1$ vertex in $[\beta, 2\beta)$. The weights of this configuration are given in Eqs.~(\ref{eq:ZAweight}-\ref{eq:Z2weight}). (b) The extended configuration space combines two ensembles $\mathcal{Z}^{2}$ and $\mathcal{Z}^{A}$. MC updates Eqs.~(\ref{eq:add}-\ref{eq:remove}) change the vertex configuration and Eq.~(\ref{eq:switch}) switch between the two ensembles.}
\label{fig:concepts}
\end{figure}
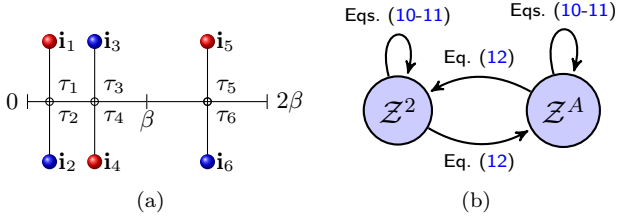

Our method is general and applicable to any fermionic system which is accessible to Monte Carlo simulations. For illustration purposes we here focus on calculating the rank-$2$ Renyi EE of an interacting spinless fermions model 

\begin{equation}
 \hat{H}= -t \sum_{\langle \mathbf{i,j}\rangle}\left(  \hat{c}_{\mathbf{i}}^{\dagger} \hat{c}_{\mathbf{j}} +  \hat{c}_{\mathbf{j}}^{\dagger} \hat{c}_{\mathbf{i}}  \right) + V \sum_{\langle \mathbf{i,j} \rangle} \left( \hat{n}_{\mathbf{i}} - \frac{1}{2}
  \right) \left( \hat{n}_{\mathbf{j}} - \frac{1}{2} \right). 
 \label{eq:Ham}
\end{equation} 
In the CTQMC method, the partition function ratios are expanded in terms of the interaction vertices~\cite{Rubtsov:2005iw, Gull:2011jd, CTQMCpaper}, 

\begin{widetext}
\begin{eqnarray}
\eta \frac{\mathcal{Z}^{A}}{\mathcal{Z}_{0}^{A}}  & =& \eta \sum_{k=0}^{\infty} 
\frac{(-V)^{k}}{k!} \int_{0}^{2\beta}\mathrm{d} \tau_{2} \int_{0}^{2\beta}\mathrm{d} \tau_{4}  \ldots \int_{0}^{2\beta} \mathrm{d}\tau_{2k}   \det\left(G^{k}_{\mathcal{Z}^{A}}\right) \label{eq:ZAoverZA0}  =  \sum_{k}\sum_{\mathcal{C}^{k}} w_{\mathcal{Z}^{A}}(\mathcal{C}^{k}), \\   
\left(\frac{\mathcal{Z}}{\mathcal{Z}_{0}}\right)^{2} & =& \sum_{k_{1}=0}^{\infty} \sum_{k_{2}=0}^{\infty} 
\frac{(-V)^{k_{1}+k_{2}}}{k_{1}!k_{2}!}  \int_{0}^{\beta} \mathrm{d}\tau_{2} \ldots \int_{0}^{\beta} \mathrm{d}\tau_{2k_{1}} \det(G_{\mathcal{Z}}^{k_{1}}) \, \int_{\beta}^{2\beta} \mathrm{d}\tau_{2k_{1}+2} \ldots \int_{\beta}^{2\beta} \mathrm{d}\tau_{2k_{1}+2k_{2}}\det{(G_{\mathcal{Z}}^{k_{2}})} \nonumber \\& =& \sum_{k_{1},k_{2}}\sum_{\mathcal{C}^{k_{1}+k_{2}}} w_{\mathcal{Z}^{2}}(\mathcal{C}^{k_{1}+k_{2}}).
\label{eq:ZoverZ0}
\end{eqnarray}
\end{widetext}
We can treat Eq.~(\ref{eq:ZAoverZA0}) and Eq.~(\ref{eq:ZoverZ0}) on equal footing and use the same configuration $\mathcal{C}^{k=k_{1}+k_{2}}$ for each term in the sampling.  Fig.~\ref{fig:concepts}(a) shows an example configuration with $k=3$ vertices, where the imaginary times satisfy $0\le\tau_{1}=\tau_{2} < \ldots < \tau_{2k_{1}-1} = \tau_{2k_{1}} < \beta \le \tau_{2k_{1}+1}= \tau_{2k_{1}+2}  \ldots < \tau_{2k-1} = \tau_{2k} < 2\beta $. Any of the configurations $\mathcal{C}^{k}$ is a valid configuration in both ensembles, but with different weights 

\begin{eqnarray}
w_{\mathcal{Z}^{A}}(\mathcal{C}^{k}) &=& \eta (-V)^{k} \det(G^{k}_{\mathcal{Z}^{A}})  \label{eq:ZAweight} \\
w_{\mathcal{Z}^{2}}(\mathcal{C}^{k}) &= & (-V)^{k} \det(G^{k_{1}}_{\mathcal{Z}}) \det(G^{k_{2}}_{\mathcal{Z}})
\label{eq:Z2weight}
\end{eqnarray}
where $G^{k}_{\mathcal{Z}^{A}}$ and $G_{\mathcal{Z}}^{k_{1(2)}}$ are $2k\times 2k$ and $2k_{1(2)}\times 2k_{1(2)}$ matrices whose matrix elements only depend on the noninteracting Green's functions~\cite{SM}. On a bipartite lattice with repulsive interaction $V>0$ the weights are positive~\cite{Huffman:2014fj, CTQMCpaper} and there is no sign problem in the Monte Carlo simulation.

We introduce an ensemble flag $\mathbb{X}\in \{\mathcal{Z}^{A}, \mathcal{Z}^{2}\}$ and perform MC simulation in an extended ensemble \cite{Burovski:2006hv, Humeniuk:2012cq} with the partition function $(\mathcal{Z}/\mathcal{Z}_{0})^{2} + \eta (\mathcal{Z}^{A}/\mathcal{Z}_{0}^{A}) = \sum_{\mathbb{X}}\sum_{k, \mathcal{C}^{k}}w_\mathbb{X} (\mathcal{C}^{k})$. Now a MC configuration corresponds to a given set of variables: the ensemble flag $\mathbb{X}$, the perturbation order $k$ and the vertex configurations. Two kinds of MC updates are necessary to ensure ergodicity of the sampling, shown in Fig.{\ref{fig:concepts}}(b). First, we keep the ensemble flag $\mathbb{X}$ unchanged and update the vertex configuration by either adding or removing one vertex. This is done either by proposing a candidate vertex at a random time (in the interval $[0, 2\beta)$) and a random bond (out of $N_{b}$ possible ones) or randomly choosing an existing vertex (out of $k$ possible ones) to be removed. The acceptance ratios are

\begin{eqnarray}
R_{\{\mathbb{X}, \mathcal{C}^{k}\}\rightarrow \{\mathbb{X}, \mathcal{C}^{k+1}\}} & =&  \frac{2\beta  N_{b}}{k+1} \frac{w_\mathbb{X}(\mathcal{C}^{k+1})}{w_\mathbb{X}(\mathcal{C}^{k})} \label{eq:add} \\
R_{\{\mathbb{X}, \mathcal{C}^{k}\}\rightarrow \{\mathbb{X}, \mathcal{C}^{k-1}\}} & =&  \frac{k}{2\beta  N_{b}} \frac{w_\mathbb{X}(\mathcal{C}^{k-1})}{w_\mathbb{X}(\mathcal{C}^{k})} \label{eq:remove}
\end{eqnarray}
The second class of MC updates switch the ensemble $\mathbb{X}$ to $\mathbb{X}^{\prime}$ while keeping the configuration $\mathcal{C}^{k}$ fixed. The acceptance probability is

\begin{eqnarray}
R_{\{\mathbb{X}, \mathcal{C}^{k}\} \rightarrow \{\mathbb{X}^{\prime}, \mathcal{C}^{k}\} } = {w_{\mathbb{X}^{\prime}}(\mathcal{C}^{k})}/{w_{\mathbb{X}}(\mathcal{C}^{k})}
\label{eq:switch}
\end{eqnarray} 
A direct evaluation of $R_{\{\mathbb{X}, \mathcal{C}^{k}\} \rightarrow \{\mathbb{X}^{\prime}, \mathcal{C}^{k}\}}$ is numerically expensive and unstable, since it requires the calculation of two determinants Eq.~(\ref{eq:ZAweight}) and Eq.~(\ref{eq:Z2weight}) and their ratio. We utilize the fact that the determinant ratio of the zeroth expansion order is one and keep updating the ratio Eq.~(\ref{eq:switch}) during the MC updates. Therefore the ensemble switch can then be implemented without any matrix operations and is very cheap. 

The relative time spent in each ensemble~\cite{Humeniuk:2012cq}  provides an estimator of the ratio between Eq.~(\ref{eq:ZAoverZA0}) and Eq.~(\ref{eq:ZoverZ0}), thus 

\begin{equation}
\Delta S_{2} = -\ln\left [\frac{\langle \delta_{\mathbb{X},\mathcal{Z}^{A}} \rangle_\mathrm{MC}}{ \langle \delta _{\mathbb{X}, \mathcal{Z}^{2}} \rangle_\mathrm{MC}} \right ] + \ln(\eta).
\label{eq:estimator}
\end{equation}

The fact that MC simulation only samples the interaction corrections of the Renyi EE $\Delta S_{2}$ is the major advantage of the present approach. On top of that, the parameter $\eta$ provides additional control over the MC dynamics. An ideal choice of $\eta$ will balance the probability of each ensemble and the observable in the square bracket of Eq.~(\ref{eq:estimator}) will be of order one. 


\begin{figure}[t]
\centering
\includegraphics[width=9cm]{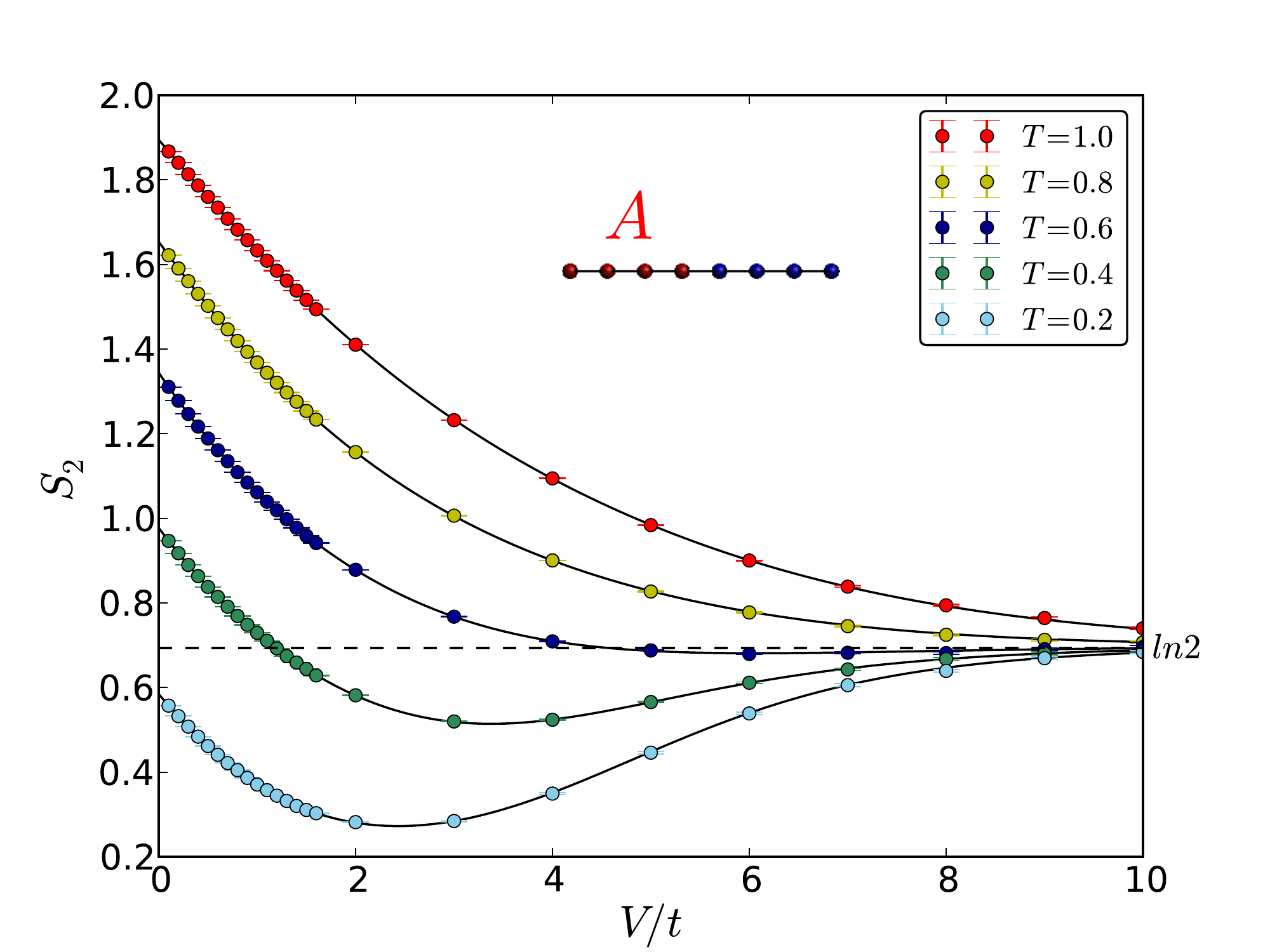}
\caption{Renyi EE of an $8$-site open chain with equal partitions. The black solid lines are the exact diagonalization results, which agree perfectly with the results calculated by our algorithm (circles).}
\label{fig:1dbenchmark}
\end{figure}

Fig.~\ref{fig:1dbenchmark} shows benchmark results of Renyi EE of an $8$-site chain with open boundaries. Our CTQMC results perfectly reproduce the exact diagonalization results (black solid lines) for all interaction strengths and temperatures. In the strong coupling limit $V\gg t $ the system has two fold quasi-degenerated ground states corresponding to the two staggered charge-density-wave (CDW) configurations. They are separated from other states by an excitation gap $\sim V$. The Renyi EE indeed approaches to $\ln 2$ when the excitation gap is much larger than the temperature.  
  

Although the above algorithm can already handle a large class of interesting problems, it will suffer from very low transition probability when the two ensembles $\mathcal{Z}^{A}$ and $\mathcal{Z}^{2}$ have vanishing small overlap in configurational space (for example if their average perturbation order differs substantially). To achieve better performance, we additionally employ the ratio method ~\cite{Hastings:2010dca, Humeniuk:2012cq} and split the partition function ratio into products of a series of intermediate ratios, each one corresponds to two ensembles where the size of region A only differs by a small amount of sites~\footnote{In the case the weight Eq.~(\ref{eq:Z2weight}) is replaced by Eq.~(\ref{eq:ZAweight}) with a different subregion A}. Because $\Delta S_{2}$ can be much smaller than  the total entanglement entropy $S_{2}$, we need far less intermediate ratios compare to the other approach~\cite{Broecker:2014ud} to achieve accurate results.

\begin{figure}[t]
\centering
\includegraphics[width=9cm]{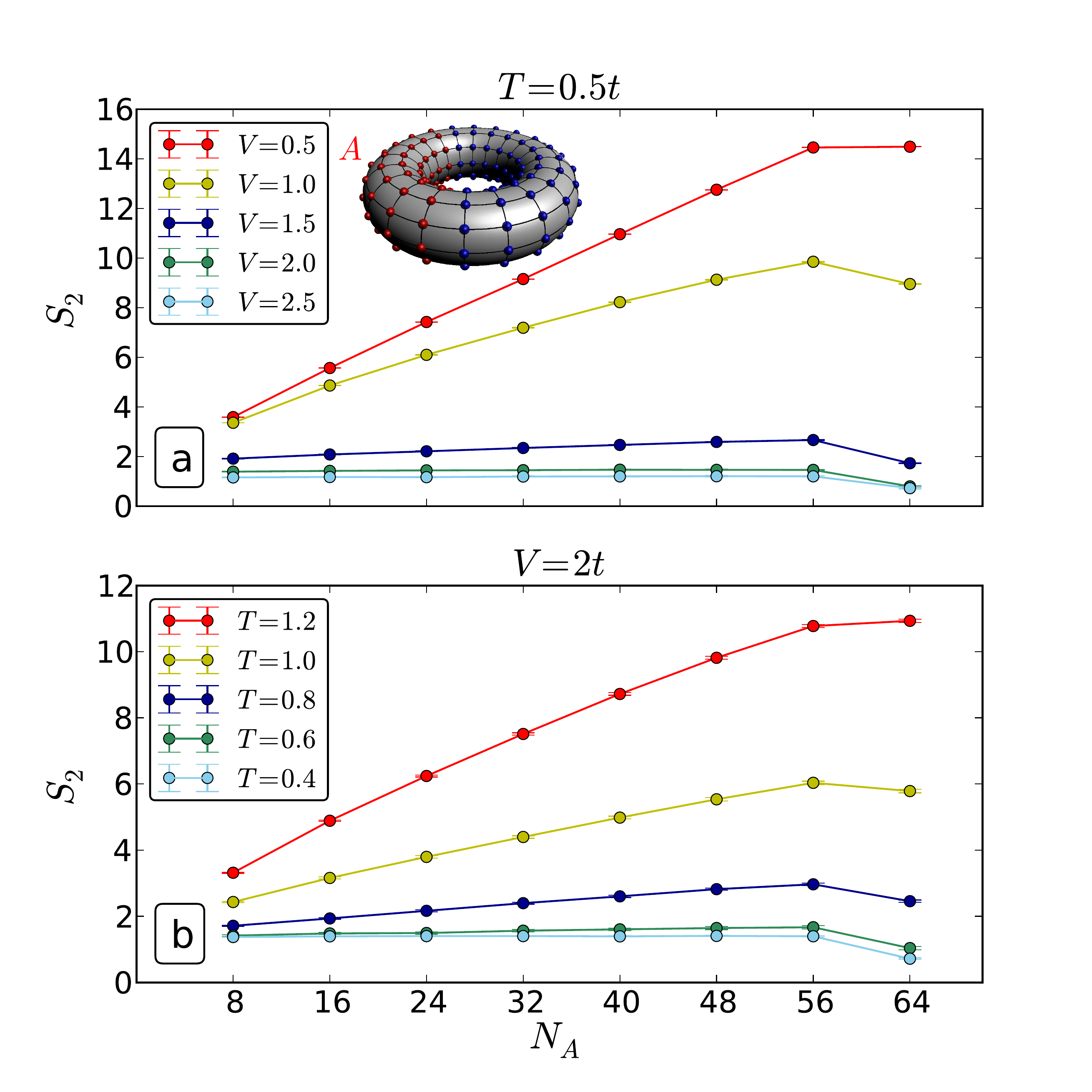}
\caption{The rank-2 Renyi entanglement entropy versus the subregion size $N_{A}$ of a cylinder embedded in a $8\times8$ torus for (a) $T=0.5t$ and (b) $V=2t$.}
\label{fig:scanNA}
\end{figure}

Our method is efficient for simulating~\emph{highly entangled} systems because it samples the difference of entanglement entropy to that of the free fermions. For obvious application like weakly interacting fermions, the algorithm enjoys additional advantage of dealing with small average expansion order, which is proportional to the interaction energy~\cite{Rubtsov:2005iw, Gull:2011jd}. 

We will demonstrate the power of our method with a more challenging example, where interaction turns the free fermions state into another phase which follows different entanglement entropy scaling. We consider the model (\ref{eq:Ham}) on a square lattice at half-filling \cite{PhysRevB.29.5253, Gubernatis:1985wo}. At zero temperate it has CDW ground state for arbitrary weak repulsive interaction because of Fermi surface nesting. The CDW ground state in the strong coupling limit can be interpreted as an  ``antiferromagnetic Ising'' ground state of a classical lattice gas. Upon increasing the temperature, the CDW state undergoes an Ising phase transition. The transition temperature is exponentially small in the weak coupling limit and is proportional to the interaction strength $V$ in the strong coupling limit.  Figure 20 of Ref.\cite{Gubernatis:1985wo} reports the $T-V$ phase diagram of this model. 

\begin{figure}[t]
\centering
\includegraphics[width=9cm]{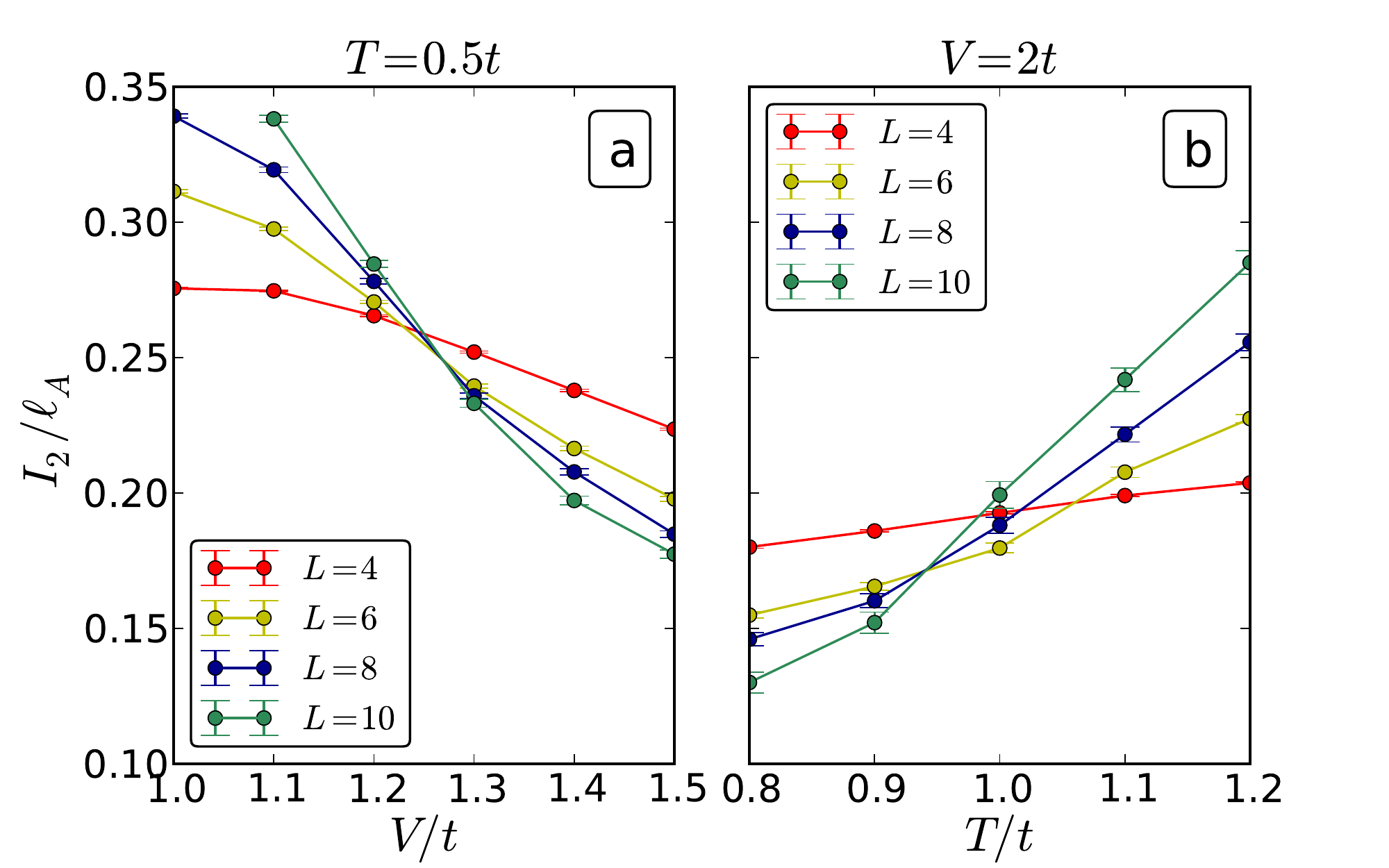}
\caption{The scaled mutual information for various system sizes at (a) $T/t=0.5$ and (b) $V/t=2$. The crossing point agrees with transition point determined in Ref.~\cite{Gubernatis:1985wo} using correlation functions.}
\label{fig:MI}
\end{figure}

We here revisit this problem from a quantum information perspective by calculating the Renyi EE across the phase transition. We consider a cylindrical subregion A embedded in a $L\times L$ torus,  Fig.~\ref{fig:scanNA}(a) inset. The boundary length is chosen to be $\ell_{A} =2L$ and is independent of the subregion size $N_{A}$. Figure~\ref{fig:scanNA} shows the rank-$2$ Renyi EE as a function $N_{A}$ at various temperature and interaction strengths, which clearly exhibits two different characteristic behaviors. At high temperature or weak interaction the Renyi EE increases linearly with $N_{A}$, indicating a highly entangled state with a volume scaling law for the EE. At low temperature or for strong interaction the Renyi EE is strongly suppressed and is essentially flat  as $N_{A}$ increases, which is a characteristic behavior of a gapful CDW state. This state is approximately a linear superposition of two simple product states. Figure~\ref{fig:scanNA}(a) also suggests that simulations deep inside the CDW state may not be ideal for the present method because the EE is much smaller than the free fermion state. The fact that we can nevertheless easily obtain correct results in this region provides a stringent test of the proposed method.

We next proceed to a quantitative determination of the phase boundaries. We consider the  mutual information~\cite{Wolf:2008hta, Melko:2010jda}
\begin{equation}
I_{2}(A:B) = S_{2}(\hat{\rho}_{A}) + S_{2}(\hat{\rho}_{B}) - S_{2}(\hat{\rho}_{A\cup B}) 
\label{eq:MI}
\end{equation} 
which cancels the bulk contribution in $S_{2}$ and exhibits boundary law even at nonzero temperature. Studies of classical and quantum spin models~\cite{Melko:2010jda, PhysRevLett.106.135701,  Iaconis:2013jz, PhysRevE.89.012125} show that the scaled mutual information $I_{2}/\ell_{A}$ crosses around the transition point. We consider a region A with $N_{A}=L^{2}/2$ and the mutual information can be directly calculated using two data points of $S_{2}$ in the Fig.\ref{fig:scanNA}. 

Figure~\ref{fig:MI} shows the scaled mutual information versus interaction strength and temperature for several system sizes. The crossing points of $I_{2}/\ell_{A}$ provide an estimation of transition point $V_{c}\in[1.2, 1.3]$ at $T=0.5t$ or $T_{c}\in[0.9, 1.0]$ at $V=2t$, which is consistent with the phase diagram reported in~\cite{Gubernatis:1985wo}. Similar to the case of Ref.~\cite{Melko:2010jda, Iaconis:2013jz}, the curves of mutual information also cross around $2T_{c}$ (not shown). These results indicate that our approach can be a versatile tool to calculate the Renyi EE and detect phase transitions in the interacting fermionic models.  


%

The proposed method allows efficient calculation of the Renyi entanglement entropy of interacting fermions by reallocating computational resources to ~\emph{interaction corrections}. Compare to the direct approaches~\cite{Grover:2013cs, Broecker:2014ud}, it can avoid sampling exponentially rare events caused by fast increase of free fermion entanglement entropy. The method is applicable to any fermionic system which could be simulated with CTQMC method and may provide further insights to the unconventional quantum critical point~\cite{PhysRevX.3.031010, CTQMCpaper} and the topological phase transitions~\cite{Hohenadler:2011kk, Wang:2014vb, Anonymous:XiXakTeu}. Our method is an ideal tool to offer an entanglement perspective on the Kondo problem~\cite{Sorensen:2007eg,Sorensen:2007dn, PhysRevB.84.041107, PhysRevLett.109.066403}, which is intimately related to the topological entanglement entropy~\cite{Levin:2006ij, Kitaev:2006dn} of two dimensional gapful states~\cite{Fendley:2007gkc}. Most interestingly, the ability to calculate entanglement entropy in CTQMC offers a portal to study entanglement in the framework of dynamical-mean-field-theory~\cite{RevModPhys.68.13,Gull:2011jd} and will  shed light on entanglement properties of realistic correlated materials. 

\paragraph{Acknowledgment}
We thank Nikolay Prokof'ev for insightful suggestions. L.W. acknowledges KITPC Beijing for hospitality during the workshop ``Precision Many-body Physics of Strongly Correlated Quantum Matter'', where part of this work was done. Simulations were performed on the M\"{o}nch cluster of Platform for Advanced Scientific Computing (PASC), the Brutus cluster at ETH Zurich and the ``Monte Rosa'' Cray XE6 at the Swiss National Supercomputing Centre (CSCS). We have used ALPS libraries~\cite{BBauer:2011tz} for Monte Carlo simulations and data analysis. This work was supported by ERC Advanced Grant SIMCOFE.

\bibliographystyle{apsrev4-1}
\bibliography{fermionEE}

\begin{thebibliography}{52}%
\makeatletter
\providecommand \@ifxundefined [1]{%
 \@ifx{#1\undefined}
}%
\providecommand \@ifnum [1]{%
 \ifnum #1\expandafter \@firstoftwo
 \else \expandafter \@secondoftwo
 \fi
}%
\providecommand \@ifx [1]{%
 \ifx #1\expandafter \@firstoftwo
 \else \expandafter \@secondoftwo
 \fi
}%
\providecommand \natexlab [1]{#1}%
\providecommand \enquote  [1]{``#1''}%
\providecommand \bibnamefont  [1]{#1}%
\providecommand \bibfnamefont [1]{#1}%
\providecommand \citenamefont [1]{#1}%
\providecommand \href@noop [0]{\@secondoftwo}%
\providecommand \href [0]{\begingroup \@sanitize@url \@href}%
\providecommand \@href[1]{\@@startlink{#1}\@@href}%
\providecommand \@@href[1]{\endgroup#1\@@endlink}%
\providecommand \@sanitize@url [0]{\catcode `\\12\catcode `\$12\catcode
  `\&12\catcode `\#12\catcode `\^12\catcode `\_12\catcode `\%12\relax}%
\providecommand \@@startlink[1]{}%
\providecommand \@@endlink[0]{}%
\providecommand \url  [0]{\begingroup\@sanitize@url \@url }%
\providecommand \@url [1]{\endgroup\@href {#1}{\urlprefix }}%
\providecommand \urlprefix  [0]{URL }%
\providecommand \Eprint [0]{\href }%
\providecommand \doibase [0]{http://dx.doi.org/}%
\providecommand \selectlanguage [0]{\@gobble}%
\providecommand \bibinfo  [0]{\@secondoftwo}%
\providecommand \bibfield  [0]{\@secondoftwo}%
\providecommand \translation [1]{[#1]}%
\providecommand \BibitemOpen [0]{}%
\providecommand \bibitemStop [0]{}%
\providecommand \bibitemNoStop [0]{.\EOS\space}%
\providecommand \EOS [0]{\spacefactor3000\relax}%
\providecommand \BibitemShut  [1]{\csname bibitem#1\endcsname}%
\let\auto@bib@innerbib\@empty
\bibitem [{\citenamefont {Feiguin}\ \emph {et~al.}(2007)\citenamefont
  {Feiguin}, \citenamefont {Trebst}, \citenamefont {Ludwig}, \citenamefont
  {Troyer}, \citenamefont {Kitaev}, \citenamefont {Wang},\ and\ \citenamefont
  {Freedman}}]{Feiguin:2007bo}%
  \BibitemOpen
  \bibfield  {author} {\bibinfo {author} {\bibfnamefont {A.}~\bibnamefont
  {Feiguin}}, \bibinfo {author} {\bibfnamefont {S.}~\bibnamefont {Trebst}},
  \bibinfo {author} {\bibfnamefont {A.}~\bibnamefont {Ludwig}}, \bibinfo
  {author} {\bibfnamefont {M.}~\bibnamefont {Troyer}}, \bibinfo {author}
  {\bibfnamefont {A.}~\bibnamefont {Kitaev}}, \bibinfo {author} {\bibfnamefont
  {Z.}~\bibnamefont {Wang}}, \ and\ \bibinfo {author} {\bibfnamefont
  {M.}~\bibnamefont {Freedman}},\ }\href {\doibase
  10.1103/PhysRevLett.98.160409} {\bibfield  {journal} {\bibinfo  {journal}
  {Phys. Rev. Lett.}\ }\textbf {\bibinfo {volume} {98}},\ \bibinfo {pages}
  {160409} (\bibinfo {year} {2007})}\BibitemShut {NoStop}%
\bibitem [{\citenamefont {Bonesteel}\ and\ \citenamefont
  {Yang}(2007)}]{PhysRevLett.99.140405}%
  \BibitemOpen
  \bibfield  {author} {\bibinfo {author} {\bibfnamefont {N.~E.}\ \bibnamefont
  {Bonesteel}}\ and\ \bibinfo {author} {\bibfnamefont {K.}~\bibnamefont
  {Yang}},\ }\href {\doibase 10.1103/PhysRevLett.99.140405} {\bibfield
  {journal} {\bibinfo  {journal} {Phys. Rev. Lett.}\ }\textbf {\bibinfo
  {volume} {99}},\ \bibinfo {pages} {140405} (\bibinfo {year}
  {2007})}\BibitemShut {NoStop}%
\bibitem [{\citenamefont {Isakov}\ \emph {et~al.}(2011)\citenamefont {Isakov},
  \citenamefont {Hastings},\ and\ \citenamefont {Melko}}]{Isakov:2011fz}%
  \BibitemOpen
  \bibfield  {author} {\bibinfo {author} {\bibfnamefont {S.~V.}\ \bibnamefont
  {Isakov}}, \bibinfo {author} {\bibfnamefont {M.~B.}\ \bibnamefont
  {Hastings}}, \ and\ \bibinfo {author} {\bibfnamefont {R.~G.}\ \bibnamefont
  {Melko}},\ }\href {http://www.nature.com/doifinder/10.1038/nphys2036}
  {\bibfield  {journal} {\bibinfo  {journal} {Nature Physics}\ }\textbf
  {\bibinfo {volume} {7}},\ \bibinfo {pages} {772} (\bibinfo {year}
  {2011})}\BibitemShut {NoStop}%
\bibitem [{\citenamefont {Jiang}\ \emph {et~al.}(2012)\citenamefont {Jiang},
  \citenamefont {Wang},\ and\ \citenamefont {Balents}}]{Jiang:2012dw}%
  \BibitemOpen
  \bibfield  {author} {\bibinfo {author} {\bibfnamefont {H.-C.}\ \bibnamefont
  {Jiang}}, \bibinfo {author} {\bibfnamefont {Z.}~\bibnamefont {Wang}}, \ and\
  \bibinfo {author} {\bibfnamefont {L.}~\bibnamefont {Balents}},\ }\href
  {http://www.nature.com/nphys/journal/v8/n12/abs/nphys2465.html} {\bibfield
  {journal} {\bibinfo  {journal} {Nature Physics}\ }\textbf {\bibinfo {volume}
  {8}},\ \bibinfo {pages} {902} (\bibinfo {year} {2012})}\BibitemShut {NoStop}%
\bibitem [{\citenamefont {Depenbrock}\ \emph {et~al.}(2012)\citenamefont
  {Depenbrock}, \citenamefont {McCulloch},\ and\ \citenamefont
  {Schollw\"ock}}]{PhysRevLett.109.067201}%
  \BibitemOpen
  \bibfield  {author} {\bibinfo {author} {\bibfnamefont {S.}~\bibnamefont
  {Depenbrock}}, \bibinfo {author} {\bibfnamefont {I.~P.}\ \bibnamefont
  {McCulloch}}, \ and\ \bibinfo {author} {\bibfnamefont {U.}~\bibnamefont
  {Schollw\"ock}},\ }\href {\doibase 10.1103/PhysRevLett.109.067201} {\bibfield
   {journal} {\bibinfo  {journal} {Phys. Rev. Lett.}\ }\textbf {\bibinfo
  {volume} {109}},\ \bibinfo {pages} {067201} (\bibinfo {year}
  {2012})}\BibitemShut {NoStop}%
\bibitem [{\citenamefont {Wen}(1990)}]{Wen:1990tm}%
  \BibitemOpen
  \bibfield  {author} {\bibinfo {author} {\bibfnamefont {X.-G.}\ \bibnamefont
  {Wen}},\ }\href
  {http://www.worldscientific.com/doi/abs/10.1142/S0217979290000139} {\bibfield
   {journal} {\bibinfo  {journal} {Int. J. Mod. Phys. B}\ }\textbf {\bibinfo
  {volume} {4}},\ \bibinfo {pages} {239} (\bibinfo {year} {1990})}\BibitemShut
  {NoStop}%
\bibitem [{\citenamefont {Buividovich}\ and\ \citenamefont
  {Polikarpov}(2008)}]{Buividovich:2008hz}%
  \BibitemOpen
  \bibfield  {author} {\bibinfo {author} {\bibfnamefont {P.~V.}\ \bibnamefont
  {Buividovich}}\ and\ \bibinfo {author} {\bibfnamefont {M.~I.}\ \bibnamefont
  {Polikarpov}},\ }\href {\doibase 10.1016/j.nuclphysb.2008.04.024} {\bibfield
  {journal} {\bibinfo  {journal} {Nuclear Physics B}\ }\textbf {\bibinfo
  {volume} {802}},\ \bibinfo {pages} {458} (\bibinfo {year}
  {2008})}\BibitemShut {NoStop}%
\bibitem [{\citenamefont {Hastings}\ \emph {et~al.}(2010)\citenamefont
  {Hastings}, \citenamefont {Gonz{\'a}lez}, \citenamefont {Kallin},\ and\
  \citenamefont {Melko}}]{Hastings:2010dca}%
  \BibitemOpen
  \bibfield  {author} {\bibinfo {author} {\bibfnamefont {M.~B.}\ \bibnamefont
  {Hastings}}, \bibinfo {author} {\bibfnamefont {I.}~\bibnamefont
  {Gonz{\'a}lez}}, \bibinfo {author} {\bibfnamefont {A.~B.}\ \bibnamefont
  {Kallin}}, \ and\ \bibinfo {author} {\bibfnamefont {R.~G.}\ \bibnamefont
  {Melko}},\ }\href {http://link.aps.org/doi/10.1103/PhysRevLett.104.157201}
  {\bibfield  {journal} {\bibinfo  {journal} {Phys. Rev. Lett.}\ }\textbf
  {\bibinfo {volume} {104}},\ \bibinfo {pages} {157201} (\bibinfo {year}
  {2010})}\BibitemShut {NoStop}%
\bibitem [{\citenamefont {Herdman}\ \emph {et~al.}(2014)\citenamefont
  {Herdman}, \citenamefont {Roy}, \citenamefont {Melko},\ and\ \citenamefont
  {Del~Maestro}}]{Herdman:2014jqa}%
  \BibitemOpen
  \bibfield  {author} {\bibinfo {author} {\bibfnamefont {C.~M.}\ \bibnamefont
  {Herdman}}, \bibinfo {author} {\bibfnamefont {P.~N.}\ \bibnamefont {Roy}},
  \bibinfo {author} {\bibfnamefont {R.~G.}\ \bibnamefont {Melko}}, \ and\
  \bibinfo {author} {\bibfnamefont {A.}~\bibnamefont {Del~Maestro}},\ }\href
  {http://link.aps.org/doi/10.1103/PhysRevB.89.140501} {\bibfield  {journal}
  {\bibinfo  {journal} {Phys. Rev. B}\ }\textbf {\bibinfo {volume} {89}},\
  \bibinfo {pages} {140501} (\bibinfo {year} {2014})}\BibitemShut {NoStop}%
\bibitem [{\citenamefont {Zhang}\ \emph {et~al.}(2011)\citenamefont {Zhang},
  \citenamefont {Grover},\ and\ \citenamefont {Vishwanath}}]{Zhang:2011ka}%
  \BibitemOpen
  \bibfield  {author} {\bibinfo {author} {\bibfnamefont {Y.}~\bibnamefont
  {Zhang}}, \bibinfo {author} {\bibfnamefont {T.}~\bibnamefont {Grover}}, \
  and\ \bibinfo {author} {\bibfnamefont {A.}~\bibnamefont {Vishwanath}},\
  }\href {http://link.aps.org/doi/10.1103/PhysRevLett.107.067202} {\bibfield
  {journal} {\bibinfo  {journal} {Phys. Rev. Lett.}\ }\textbf {\bibinfo
  {volume} {107}},\ \bibinfo {pages} {067202} (\bibinfo {year}
  {2011})}\BibitemShut {NoStop}%
\bibitem [{\citenamefont {McMinis}\ and\ \citenamefont
  {Tubman}(2013)}]{McMinis:2013dp}%
  \BibitemOpen
  \bibfield  {author} {\bibinfo {author} {\bibfnamefont {J.}~\bibnamefont
  {McMinis}}\ and\ \bibinfo {author} {\bibfnamefont {N.~M.}\ \bibnamefont
  {Tubman}},\ }\href {http://link.aps.org/doi/10.1103/PhysRevB.87.081108}
  {\bibfield  {journal} {\bibinfo  {journal} {Phys. Rev. B}\ }\textbf {\bibinfo
  {volume} {87}},\ \bibinfo {pages} {081108} (\bibinfo {year}
  {2013})}\BibitemShut {NoStop}%
\bibitem [{\citenamefont {Grover}(2013)}]{Grover:2013cs}%
  \BibitemOpen
  \bibfield  {author} {\bibinfo {author} {\bibfnamefont {T.}~\bibnamefont
  {Grover}},\ }\href {http://link.aps.org/doi/10.1103/PhysRevLett.111.130402}
  {\bibfield  {journal} {\bibinfo  {journal} {Phys. Rev. Lett.}\ }\textbf
  {\bibinfo {volume} {111}},\ \bibinfo {pages} {130402} (\bibinfo {year}
  {2013})}\BibitemShut {NoStop}%
\bibitem [{\citenamefont {Broecker}\ and\ \citenamefont
  {Trebst}(2014)}]{Broecker:2014ud}%
  \BibitemOpen
  \bibfield  {author} {\bibinfo {author} {\bibfnamefont {P.}~\bibnamefont
  {Broecker}}\ and\ \bibinfo {author} {\bibfnamefont {S.}~\bibnamefont
  {Trebst}},\ }\href {http://arxiv.org/abs/1404.3027} {\bibfield  {journal}
  {\bibinfo  {journal} {arXiv:1404.3027}\ } (\bibinfo {year}
  {2014})}\BibitemShut {NoStop}%
\bibitem [{\citenamefont {Levin}\ and\ \citenamefont
  {Wen}(2006)}]{Levin:2006ij}%
  \BibitemOpen
  \bibfield  {author} {\bibinfo {author} {\bibfnamefont {M.}~\bibnamefont
  {Levin}}\ and\ \bibinfo {author} {\bibfnamefont {X.-G.}\ \bibnamefont
  {Wen}},\ }\href {http://link.aps.org/doi/10.1103/PhysRevLett.96.110405}
  {\bibfield  {journal} {\bibinfo  {journal} {Phys. Rev. Lett.}\ }\textbf
  {\bibinfo {volume} {96}},\ \bibinfo {pages} {110405} (\bibinfo {year}
  {2006})}\BibitemShut {NoStop}%
\bibitem [{\citenamefont {Kitaev}\ and\ \citenamefont
  {Preskill}(2006)}]{Kitaev:2006dn}%
  \BibitemOpen
  \bibfield  {author} {\bibinfo {author} {\bibfnamefont {A.}~\bibnamefont
  {Kitaev}}\ and\ \bibinfo {author} {\bibfnamefont {J.}~\bibnamefont
  {Preskill}},\ }\href {http://link.aps.org/doi/10.1103/PhysRevLett.96.110404}
  {\bibfield  {journal} {\bibinfo  {journal} {Phys. Rev. Lett.}\ }\textbf
  {\bibinfo {volume} {96}},\ \bibinfo {pages} {110404} (\bibinfo {year}
  {2006})}\BibitemShut {NoStop}%
\bibitem [{\citenamefont {Assaad}\ \emph {et~al.}(2014)\citenamefont {Assaad},
  \citenamefont {Lang},\ and\ \citenamefont
  {Parisen~Toldin}}]{2014PhRvB..89l5121A}%
  \BibitemOpen
  \bibfield  {author} {\bibinfo {author} {\bibfnamefont {F.~F.}\ \bibnamefont
  {Assaad}}, \bibinfo {author} {\bibfnamefont {T.~C.}\ \bibnamefont {Lang}}, \
  and\ \bibinfo {author} {\bibfnamefont {F.}~\bibnamefont {Parisen~Toldin}},\
  }\href
  {http://adsabs.harvard.edu/cgi-bin/nph-data_query?bibcode=2014PhRvB..89l5121A&link_type=ABSTRACT}
  {\bibfield  {journal} {\bibinfo  {journal} {Phys. Rev. B}\ }\textbf {\bibinfo
  {volume} {89}},\ \bibinfo {pages} {125121} (\bibinfo {year}
  {2014})}\BibitemShut {NoStop}%
\bibitem [{\citenamefont {Wang}\ \emph
  {et~al.}(2014{\natexlab{a}})\citenamefont {Wang}, \citenamefont {Xu},
  \citenamefont {Wang},\ and\ \citenamefont {Wu}}]{Anonymous:XiXakTeu}%
  \BibitemOpen
  \bibfield  {author} {\bibinfo {author} {\bibfnamefont {D.}~\bibnamefont
  {Wang}}, \bibinfo {author} {\bibfnamefont {S.}~\bibnamefont {Xu}}, \bibinfo
  {author} {\bibfnamefont {Y.}~\bibnamefont {Wang}}, \ and\ \bibinfo {author}
  {\bibfnamefont {C.}~\bibnamefont {Wu}},\ }\href
  {http://arxiv.org/abs/1405.2043v1} {\bibfield  {journal} {\bibinfo  {journal}
  {arXiv:1405.2043}\ } (\bibinfo {year} {2014}{\natexlab{a}})}\BibitemShut
  {NoStop}%
\bibitem [{\citenamefont {Eisert}\ \emph {et~al.}(2010)\citenamefont {Eisert},
  \citenamefont {Cramer},\ and\ \citenamefont {Plenio}}]{Eisert:2010hq}%
  \BibitemOpen
  \bibfield  {author} {\bibinfo {author} {\bibfnamefont {J.}~\bibnamefont
  {Eisert}}, \bibinfo {author} {\bibfnamefont {M.}~\bibnamefont {Cramer}}, \
  and\ \bibinfo {author} {\bibfnamefont {M.~B.}\ \bibnamefont {Plenio}},\
  }\href {http://link.aps.org/doi/10.1103/RevModPhys.82.277} {\bibfield
  {journal} {\bibinfo  {journal} {Rev. Mod. Phys.}\ }\textbf {\bibinfo {volume}
  {82}},\ \bibinfo {pages} {277} (\bibinfo {year} {2010})}\BibitemShut
  {NoStop}%
\bibitem [{\citenamefont {Wolf}(2006)}]{PhysRevLett.96.010404}%
  \BibitemOpen
  \bibfield  {author} {\bibinfo {author} {\bibfnamefont {M.~M.}\ \bibnamefont
  {Wolf}},\ }\href {\doibase 10.1103/PhysRevLett.96.010404} {\bibfield
  {journal} {\bibinfo  {journal} {Phys. Rev. Lett.}\ }\textbf {\bibinfo
  {volume} {96}},\ \bibinfo {pages} {010404} (\bibinfo {year}
  {2006})}\BibitemShut {NoStop}%
\bibitem [{\citenamefont {Gioev}\ and\ \citenamefont
  {Klich}(2006)}]{PhysRevLett.96.100503}%
  \BibitemOpen
  \bibfield  {author} {\bibinfo {author} {\bibfnamefont {D.}~\bibnamefont
  {Gioev}}\ and\ \bibinfo {author} {\bibfnamefont {I.}~\bibnamefont {Klich}},\
  }\href {\doibase 10.1103/PhysRevLett.96.100503} {\bibfield  {journal}
  {\bibinfo  {journal} {Phys. Rev. Lett.}\ }\textbf {\bibinfo {volume} {96}},\
  \bibinfo {pages} {100503} (\bibinfo {year} {2006})}\BibitemShut {NoStop}%
\bibitem [{\citenamefont {Swingle}(2010)}]{PhysRevLett.105.050502}%
  \BibitemOpen
  \bibfield  {author} {\bibinfo {author} {\bibfnamefont {B.}~\bibnamefont
  {Swingle}},\ }\href {\doibase 10.1103/PhysRevLett.105.050502} {\bibfield
  {journal} {\bibinfo  {journal} {Phys. Rev. Lett.}\ }\textbf {\bibinfo
  {volume} {105}},\ \bibinfo {pages} {050502} (\bibinfo {year}
  {2010})}\BibitemShut {NoStop}%
\bibitem [{\citenamefont {Ding}\ \emph {et~al.}(2012)\citenamefont {Ding},
  \citenamefont {Seidel},\ and\ \citenamefont {Yang}}]{PhysRevX.2.011012}%
  \BibitemOpen
  \bibfield  {author} {\bibinfo {author} {\bibfnamefont {W.}~\bibnamefont
  {Ding}}, \bibinfo {author} {\bibfnamefont {A.}~\bibnamefont {Seidel}}, \ and\
  \bibinfo {author} {\bibfnamefont {K.}~\bibnamefont {Yang}},\ }\href {\doibase
  10.1103/PhysRevX.2.011012} {\bibfield  {journal} {\bibinfo  {journal} {Phys.
  Rev. X}\ }\textbf {\bibinfo {volume} {2}},\ \bibinfo {pages} {011012}
  (\bibinfo {year} {2012})}\BibitemShut {NoStop}%
\bibitem [{\citenamefont {Lai}\ \emph {et~al.}(2013)\citenamefont {Lai},
  \citenamefont {Yang},\ and\ \citenamefont
  {Bonesteel}}]{PhysRevLett.111.210402}%
  \BibitemOpen
  \bibfield  {author} {\bibinfo {author} {\bibfnamefont {H.-H.}\ \bibnamefont
  {Lai}}, \bibinfo {author} {\bibfnamefont {K.}~\bibnamefont {Yang}}, \ and\
  \bibinfo {author} {\bibfnamefont {N.~E.}\ \bibnamefont {Bonesteel}},\ }\href
  {\doibase 10.1103/PhysRevLett.111.210402} {\bibfield  {journal} {\bibinfo
  {journal} {Phys. Rev. Lett.}\ }\textbf {\bibinfo {volume} {111}},\ \bibinfo
  {pages} {210402} (\bibinfo {year} {2013})}\BibitemShut {NoStop}%
\bibitem [{\citenamefont {Leschke}\ \emph {et~al.}(2014)\citenamefont
  {Leschke}, \citenamefont {Sobolev},\ and\ \citenamefont
  {Spitzer}}]{PhysRevLett.112.160403}%
  \BibitemOpen
  \bibfield  {author} {\bibinfo {author} {\bibfnamefont {H.}~\bibnamefont
  {Leschke}}, \bibinfo {author} {\bibfnamefont {A.~V.}\ \bibnamefont
  {Sobolev}}, \ and\ \bibinfo {author} {\bibfnamefont {W.}~\bibnamefont
  {Spitzer}},\ }\href {\doibase 10.1103/PhysRevLett.112.160403} {\bibfield
  {journal} {\bibinfo  {journal} {Phys. Rev. Lett.}\ }\textbf {\bibinfo
  {volume} {112}},\ \bibinfo {pages} {160403} (\bibinfo {year}
  {2014})}\BibitemShut {NoStop}%
\bibitem [{\citenamefont {Humeniuk}\ and\ \citenamefont
  {Roscilde}(2012)}]{Humeniuk:2012cq}%
  \BibitemOpen
  \bibfield  {author} {\bibinfo {author} {\bibfnamefont {S.}~\bibnamefont
  {Humeniuk}}\ and\ \bibinfo {author} {\bibfnamefont {T.}~\bibnamefont
  {Roscilde}},\ }\href {http://link.aps.org/doi/10.1103/PhysRevB.86.235116}
  {\bibfield  {journal} {\bibinfo  {journal} {Phys. Rev. B}\ }\textbf {\bibinfo
  {volume} {86}},\ \bibinfo {pages} {235116} (\bibinfo {year}
  {2012})}\BibitemShut {NoStop}%
\bibitem [{\citenamefont {Rubtsov}\ \emph {et~al.}(2005)\citenamefont
  {Rubtsov}, \citenamefont {Savkin},\ and\ \citenamefont
  {Lichtenstein}}]{Rubtsov:2005iw}%
  \BibitemOpen
  \bibfield  {author} {\bibinfo {author} {\bibfnamefont {A.}~\bibnamefont
  {Rubtsov}}, \bibinfo {author} {\bibfnamefont {V.}~\bibnamefont {Savkin}}, \
  and\ \bibinfo {author} {\bibfnamefont {A.}~\bibnamefont {Lichtenstein}},\
  }\href {http://link.aps.org/doi/10.1103/PhysRevB.72.035122} {\bibfield
  {journal} {\bibinfo  {journal} {Phys. Rev. B}\ }\textbf {\bibinfo {volume}
  {72}},\ \bibinfo {pages} {035122} (\bibinfo {year} {2005})}\BibitemShut
  {NoStop}%
\bibitem [{\citenamefont {Gull}\ \emph {et~al.}(2011)\citenamefont {Gull},
  \citenamefont {Millis}, \citenamefont {Lichtenstein}, \citenamefont
  {Rubtsov}, \citenamefont {Troyer},\ and\ \citenamefont
  {Werner}}]{Gull:2011jd}%
  \BibitemOpen
  \bibfield  {author} {\bibinfo {author} {\bibfnamefont {E.}~\bibnamefont
  {Gull}}, \bibinfo {author} {\bibfnamefont {A.~J.}\ \bibnamefont {Millis}},
  \bibinfo {author} {\bibfnamefont {A.~I.}\ \bibnamefont {Lichtenstein}},
  \bibinfo {author} {\bibfnamefont {A.~N.}\ \bibnamefont {Rubtsov}}, \bibinfo
  {author} {\bibfnamefont {M.}~\bibnamefont {Troyer}}, \ and\ \bibinfo {author}
  {\bibfnamefont {P.}~\bibnamefont {Werner}},\ }\href
  {http://link.aps.org/doi/10.1103/RevModPhys.83.349} {\bibfield  {journal}
  {\bibinfo  {journal} {Rev. Mod. Phys.}\ }\textbf {\bibinfo {volume} {83}},\
  \bibinfo {pages} {349} (\bibinfo {year} {2011})}\BibitemShut {NoStop}%
\bibitem [{\citenamefont {Calabrese}\ and\ \citenamefont
  {Cardy}(2004)}]{Calabrese:2004hl}%
  \BibitemOpen
  \bibfield  {author} {\bibinfo {author} {\bibfnamefont {P.}~\bibnamefont
  {Calabrese}}\ and\ \bibinfo {author} {\bibfnamefont {J.}~\bibnamefont
  {Cardy}},\ }\href
  {http://stacks.iop.org/1742-5468/2004/i=06/a=P06002?key=crossref.2fead8410474dca993bc18682138a41f}
  {\bibfield  {journal} {\bibinfo  {journal} {J. Stat. Mech.: Theor. Exp.}\
  }\textbf {\bibinfo {volume} {2004}},\ \bibinfo {pages} {P06002} (\bibinfo
  {year} {2004})}\BibitemShut {NoStop}%
\bibitem [{\citenamefont {Melko}\ \emph {et~al.}(2010)\citenamefont {Melko},
  \citenamefont {Kallin},\ and\ \citenamefont {Hastings}}]{Melko:2010jda}%
  \BibitemOpen
  \bibfield  {author} {\bibinfo {author} {\bibfnamefont {R.~G.}\ \bibnamefont
  {Melko}}, \bibinfo {author} {\bibfnamefont {A.~B.}\ \bibnamefont {Kallin}}, \
  and\ \bibinfo {author} {\bibfnamefont {M.~B.}\ \bibnamefont {Hastings}},\
  }\href {http://link.aps.org/doi/10.1103/PhysRevB.82.100409} {\bibfield
  {journal} {\bibinfo  {journal} {Phys. Rev. B}\ }\textbf {\bibinfo {volume}
  {82}},\ \bibinfo {pages} {100409} (\bibinfo {year} {2010})}\BibitemShut
  {NoStop}%
\bibitem [{\citenamefont {Peschel}(2003)}]{Peschel:2002gz}%
  \BibitemOpen
  \bibfield  {author} {\bibinfo {author} {\bibfnamefont {I.}~\bibnamefont
  {Peschel}},\ }\href
  {http://stacks.iop.org/0305-4470/36/i=14/a=101?key=crossref.1975f23741354d5a3752bc280b02b1fe}
  {\bibfield  {journal} {\bibinfo  {journal} {J. Phys. A: Math. Gen.}\ }\textbf
  {\bibinfo {volume} {36}},\ \bibinfo {pages} {L205} (\bibinfo {year}
  {2003})}\BibitemShut {NoStop}%
\bibitem [{SM()}]{SM}%
  \BibitemOpen
  \href@noop {} {\bibinfo  {journal} {See Supplemental Material for details of
  calculating noninteracting Renyi entanglement entropy and noninteracting
  Green's functions}\ }\BibitemShut {NoStop}%
\bibitem [{\citenamefont {Wang}\ \emph
  {et~al.}(2014{\natexlab{b}})\citenamefont {Wang}, \citenamefont {Corboz},\
  and\ \citenamefont {Troyer}}]{CTQMCpaper}%
  \BibitemOpen
\bibfield  {journal} {  }\bibfield  {author} {\bibinfo {author} {\bibfnamefont
  {L.}~\bibnamefont {Wang}}, \bibinfo {author} {\bibfnamefont {P.}~\bibnamefont
  {Corboz}}, \ and\ \bibinfo {author} {\bibfnamefont {M.}~\bibnamefont
  {Troyer}},\ }\href {http://arxiv.org/abs/1407.0029} {\bibfield  {journal}
  {\bibinfo  {journal} {arXiv:1407.0029}\ } (\bibinfo {year}
  {2014}{\natexlab{b}})}\BibitemShut {NoStop}%
\bibitem [{\citenamefont {Huffman}\ and\ \citenamefont
  {Chandrasekharan}(2014)}]{Huffman:2014fj}%
  \BibitemOpen
  \bibfield  {author} {\bibinfo {author} {\bibfnamefont {E.~F.}\ \bibnamefont
  {Huffman}}\ and\ \bibinfo {author} {\bibfnamefont {S.}~\bibnamefont
  {Chandrasekharan}},\ }\href
  {http://link.aps.org/doi/10.1103/PhysRevB.89.111101} {\bibfield  {journal}
  {\bibinfo  {journal} {Phys. Rev. B}\ }\textbf {\bibinfo {volume} {89}},\
  \bibinfo {pages} {111101} (\bibinfo {year} {2014})}\BibitemShut {NoStop}%
\bibitem [{\citenamefont {Burovski}\ \emph {et~al.}(2006)\citenamefont
  {Burovski}, \citenamefont {Prokof'ev}, \citenamefont {Svistunov},\ and\
  \citenamefont {Troyer}}]{Burovski:2006hv}%
  \BibitemOpen
  \bibfield  {author} {\bibinfo {author} {\bibfnamefont {E.}~\bibnamefont
  {Burovski}}, \bibinfo {author} {\bibfnamefont {N.}~\bibnamefont {Prokof'ev}},
  \bibinfo {author} {\bibfnamefont {B.}~\bibnamefont {Svistunov}}, \ and\
  \bibinfo {author} {\bibfnamefont {M.}~\bibnamefont {Troyer}},\ }\href
  {http://stacks.iop.org/1367-2630/8/i=8/a=153?key=crossref.60a86faad17c96899fdec0e642e6d830}
  {\bibfield  {journal} {\bibinfo  {journal} {New J. Phys.}\ }\textbf {\bibinfo
  {volume} {8}},\ \bibinfo {pages} {153} (\bibinfo {year} {2006})}\BibitemShut
  {NoStop}%
\bibitem [{\citenamefont {Scalapino}\ \emph {et~al.}(1984)\citenamefont
  {Scalapino}, \citenamefont {Sugar},\ and\ \citenamefont
  {Toussaint}}]{PhysRevB.29.5253}%
  \BibitemOpen
  \bibfield  {author} {\bibinfo {author} {\bibfnamefont {D.~J.}\ \bibnamefont
  {Scalapino}}, \bibinfo {author} {\bibfnamefont {R.~L.}\ \bibnamefont
  {Sugar}}, \ and\ \bibinfo {author} {\bibfnamefont {W.~D.}\ \bibnamefont
  {Toussaint}},\ }\href {\doibase 10.1103/PhysRevB.29.5253} {\bibfield
  {journal} {\bibinfo  {journal} {Phys. Rev. B}\ }\textbf {\bibinfo {volume}
  {29}},\ \bibinfo {pages} {5253} (\bibinfo {year} {1984})}\BibitemShut
  {NoStop}%
\bibitem [{\citenamefont {Gubernatis}\ \emph {et~al.}(1985)\citenamefont
  {Gubernatis}, \citenamefont {Scalapino}, \citenamefont {Sugar},\ and\
  \citenamefont {Toussaint}}]{Gubernatis:1985wo}%
  \BibitemOpen
  \bibfield  {author} {\bibinfo {author} {\bibfnamefont {J.~E.}\ \bibnamefont
  {Gubernatis}}, \bibinfo {author} {\bibfnamefont {D.~J.}\ \bibnamefont
  {Scalapino}}, \bibinfo {author} {\bibfnamefont {R.~L.}\ \bibnamefont
  {Sugar}}, \ and\ \bibinfo {author} {\bibfnamefont {W.~D.}\ \bibnamefont
  {Toussaint}},\ }\href {http://prb.aps.org/abstract/PRB/v32/i1/p103_1}
  {\bibfield  {journal} {\bibinfo  {journal} {Phys. Rev. B}\ }\textbf {\bibinfo
  {volume} {32}},\ \bibinfo {pages} {103} (\bibinfo {year} {1985})}\BibitemShut
  {NoStop}%
\bibitem [{\citenamefont {Wolf}\ \emph {et~al.}(2008)\citenamefont {Wolf},
  \citenamefont {Verstraete}, \citenamefont {Hastings},\ and\ \citenamefont
  {Cirac}}]{Wolf:2008hta}%
  \BibitemOpen
  \bibfield  {author} {\bibinfo {author} {\bibfnamefont {M.}~\bibnamefont
  {Wolf}}, \bibinfo {author} {\bibfnamefont {F.}~\bibnamefont {Verstraete}},
  \bibinfo {author} {\bibfnamefont {M.}~\bibnamefont {Hastings}}, \ and\
  \bibinfo {author} {\bibfnamefont {J.}~\bibnamefont {Cirac}},\ }\href
  {http://link.aps.org/doi/10.1103/PhysRevLett.100.070502} {\bibfield
  {journal} {\bibinfo  {journal} {Phys. Rev. Lett.}\ }\textbf {\bibinfo
  {volume} {100}},\ \bibinfo {pages} {070502} (\bibinfo {year}
  {2008})}\BibitemShut {NoStop}%
\bibitem [{\citenamefont {Singh}\ \emph {et~al.}(2011)\citenamefont {Singh},
  \citenamefont {Hastings}, \citenamefont {Kallin},\ and\ \citenamefont
  {Melko}}]{PhysRevLett.106.135701}%
  \BibitemOpen
  \bibfield  {author} {\bibinfo {author} {\bibfnamefont {R.~R.~P.}\
  \bibnamefont {Singh}}, \bibinfo {author} {\bibfnamefont {M.~B.}\ \bibnamefont
  {Hastings}}, \bibinfo {author} {\bibfnamefont {A.~B.}\ \bibnamefont
  {Kallin}}, \ and\ \bibinfo {author} {\bibfnamefont {R.~G.}\ \bibnamefont
  {Melko}},\ }\href {\doibase 10.1103/PhysRevLett.106.135701} {\bibfield
  {journal} {\bibinfo  {journal} {Phys. Rev. Lett.}\ }\textbf {\bibinfo
  {volume} {106}},\ \bibinfo {pages} {135701} (\bibinfo {year}
  {2011})}\BibitemShut {NoStop}%
\bibitem [{\citenamefont {Iaconis}\ \emph {et~al.}(2013)\citenamefont
  {Iaconis}, \citenamefont {Inglis}, \citenamefont {Kallin},\ and\
  \citenamefont {Melko}}]{Iaconis:2013jz}%
  \BibitemOpen
  \bibfield  {author} {\bibinfo {author} {\bibfnamefont {J.}~\bibnamefont
  {Iaconis}}, \bibinfo {author} {\bibfnamefont {S.}~\bibnamefont {Inglis}},
  \bibinfo {author} {\bibfnamefont {A.~B.}\ \bibnamefont {Kallin}}, \ and\
  \bibinfo {author} {\bibfnamefont {R.~G.}\ \bibnamefont {Melko}},\ }\href
  {http://link.aps.org/doi/10.1103/PhysRevB.87.195134} {\bibfield  {journal}
  {\bibinfo  {journal} {Phys. Rev. B}\ }\textbf {\bibinfo {volume} {87}},\
  \bibinfo {pages} {195134} (\bibinfo {year} {2013})}\BibitemShut {NoStop}%
\bibitem [{\citenamefont {Storms}\ and\ \citenamefont
  {Singh}(2014)}]{PhysRevE.89.012125}%
  \BibitemOpen
  \bibfield  {author} {\bibinfo {author} {\bibfnamefont {M.}~\bibnamefont
  {Storms}}\ and\ \bibinfo {author} {\bibfnamefont {R.~R.~P.}\ \bibnamefont
  {Singh}},\ }\href {\doibase 10.1103/PhysRevE.89.012125} {\bibfield  {journal}
  {\bibinfo  {journal} {Phys. Rev. E}\ }\textbf {\bibinfo {volume} {89}},\
  \bibinfo {pages} {012125} (\bibinfo {year} {2014})}\BibitemShut {NoStop}%
\bibitem [{\citenamefont {Assaad}\ and\ \citenamefont
  {Herbut}(2013)}]{PhysRevX.3.031010}%
  \BibitemOpen
  \bibfield  {author} {\bibinfo {author} {\bibfnamefont {F.~F.}\ \bibnamefont
  {Assaad}}\ and\ \bibinfo {author} {\bibfnamefont {I.~F.}\ \bibnamefont
  {Herbut}},\ }\href {\doibase 10.1103/PhysRevX.3.031010} {\bibfield  {journal}
  {\bibinfo  {journal} {Phys. Rev. X}\ }\textbf {\bibinfo {volume} {3}},\
  \bibinfo {pages} {031010} (\bibinfo {year} {2013})}\BibitemShut {NoStop}%
\bibitem [{\citenamefont {Hohenadler}\ \emph {et~al.}(2011)\citenamefont
  {Hohenadler}, \citenamefont {Lang},\ and\ \citenamefont
  {Assaad}}]{Hohenadler:2011kk}%
  \BibitemOpen
  \bibfield  {author} {\bibinfo {author} {\bibfnamefont {M.}~\bibnamefont
  {Hohenadler}}, \bibinfo {author} {\bibfnamefont {T.~C.}\ \bibnamefont
  {Lang}}, \ and\ \bibinfo {author} {\bibfnamefont {F.~F.}\ \bibnamefont
  {Assaad}},\ }\href {http://link.aps.org/doi/10.1103/PhysRevLett.106.100403}
  {\bibfield  {journal} {\bibinfo  {journal} {Phys. Rev. Lett.}\ }\textbf
  {\bibinfo {volume} {106}},\ \bibinfo {pages} {100403} (\bibinfo {year}
  {2011})}\BibitemShut {NoStop}%
\bibitem [{\citenamefont {Wang}\ \emph
  {et~al.}(2014{\natexlab{c}})\citenamefont {Wang}, \citenamefont {Hung},\ and\
  \citenamefont {Troyer}}]{Wang:2014vb}%
  \BibitemOpen
  \bibfield  {author} {\bibinfo {author} {\bibfnamefont {L.}~\bibnamefont
  {Wang}}, \bibinfo {author} {\bibfnamefont {H.-H.}\ \bibnamefont {Hung}}, \
  and\ \bibinfo {author} {\bibfnamefont {M.}~\bibnamefont {Troyer}},\ }\href
  {http://arxiv.org/abs/1402.6704v1} {\bibfield  {journal} {\bibinfo  {journal}
  {arXiv:1402.6704}\ } (\bibinfo {year} {2014}{\natexlab{c}})}\BibitemShut
  {NoStop}%
\bibitem [{\citenamefont {S{\o}rensen}\ \emph
  {et~al.}(2007{\natexlab{a}})\citenamefont {S{\o}rensen}, \citenamefont
  {Chang}, \citenamefont {Laflorencie},\ and\ \citenamefont
  {Affleck}}]{Sorensen:2007eg}%
  \BibitemOpen
  \bibfield  {author} {\bibinfo {author} {\bibfnamefont {E.~S.}\ \bibnamefont
  {S{\o}rensen}}, \bibinfo {author} {\bibfnamefont {M.-S.}\ \bibnamefont
  {Chang}}, \bibinfo {author} {\bibfnamefont {N.}~\bibnamefont {Laflorencie}},
  \ and\ \bibinfo {author} {\bibfnamefont {I.}~\bibnamefont {Affleck}},\ }\href
  {\doibase 10.1088/1742-5468/2007/01/L01001} {\bibfield  {journal} {\bibinfo
  {journal} {J. Stat. Mech.: Theor. Exp.}\ }\textbf {\bibinfo {volume}
  {2007}},\ \bibinfo {pages} {L01001} (\bibinfo {year}
  {2007}{\natexlab{a}})}\BibitemShut {NoStop}%
\bibitem [{\citenamefont {S{\o}rensen}\ \emph
  {et~al.}(2007{\natexlab{b}})\citenamefont {S{\o}rensen}, \citenamefont
  {Chang}, \citenamefont {Laflorencie},\ and\ \citenamefont
  {Affleck}}]{Sorensen:2007dn}%
  \BibitemOpen
  \bibfield  {author} {\bibinfo {author} {\bibfnamefont {E.~S.}\ \bibnamefont
  {S{\o}rensen}}, \bibinfo {author} {\bibfnamefont {M.-S.}\ \bibnamefont
  {Chang}}, \bibinfo {author} {\bibfnamefont {N.}~\bibnamefont {Laflorencie}},
  \ and\ \bibinfo {author} {\bibfnamefont {I.}~\bibnamefont {Affleck}},\ }\href
  {\doibase 10.1088/1742-5468/2007/08/P08003} {\bibfield  {journal} {\bibinfo
  {journal} {J. Stat. Mech.: Theor. Exp.}\ }\textbf {\bibinfo {volume}
  {2007}},\ \bibinfo {pages} {P08003} (\bibinfo {year}
  {2007}{\natexlab{b}})}\BibitemShut {NoStop}%
\bibitem [{\citenamefont {Eriksson}\ and\ \citenamefont
  {Johannesson}(2011)}]{PhysRevB.84.041107}%
  \BibitemOpen
  \bibfield  {author} {\bibinfo {author} {\bibfnamefont {E.}~\bibnamefont
  {Eriksson}}\ and\ \bibinfo {author} {\bibfnamefont {H.}~\bibnamefont
  {Johannesson}},\ }\href {\doibase 10.1103/PhysRevB.84.041107} {\bibfield
  {journal} {\bibinfo  {journal} {Phys. Rev. B}\ }\textbf {\bibinfo {volume}
  {84}},\ \bibinfo {pages} {041107} (\bibinfo {year} {2011})}\BibitemShut
  {NoStop}%
\bibitem [{\citenamefont {Bayat}\ \emph {et~al.}(2012)\citenamefont {Bayat},
  \citenamefont {Bose}, \citenamefont {Sodano},\ and\ \citenamefont
  {Johannesson}}]{PhysRevLett.109.066403}%
  \BibitemOpen
  \bibfield  {author} {\bibinfo {author} {\bibfnamefont {A.}~\bibnamefont
  {Bayat}}, \bibinfo {author} {\bibfnamefont {S.}~\bibnamefont {Bose}},
  \bibinfo {author} {\bibfnamefont {P.}~\bibnamefont {Sodano}}, \ and\ \bibinfo
  {author} {\bibfnamefont {H.}~\bibnamefont {Johannesson}},\ }\href {\doibase
  10.1103/PhysRevLett.109.066403} {\bibfield  {journal} {\bibinfo  {journal}
  {Phys. Rev. Lett.}\ }\textbf {\bibinfo {volume} {109}},\ \bibinfo {pages}
  {066403} (\bibinfo {year} {2012})}\BibitemShut {NoStop}%
\bibitem [{\citenamefont {Fendley}\ \emph {et~al.}(2007)\citenamefont
  {Fendley}, \citenamefont {Fisher},\ and\ \citenamefont
  {Nayak}}]{Fendley:2007gkc}%
  \BibitemOpen
  \bibfield  {author} {\bibinfo {author} {\bibfnamefont {P.}~\bibnamefont
  {Fendley}}, \bibinfo {author} {\bibfnamefont {M.~P.~A.}\ \bibnamefont
  {Fisher}}, \ and\ \bibinfo {author} {\bibfnamefont {C.}~\bibnamefont
  {Nayak}},\ }\href {\doibase 10.1007/s10955-006-9275-8} {\bibfield  {journal}
  {\bibinfo  {journal} {J Stat Phys}\ }\textbf {\bibinfo {volume} {126}},\
  \bibinfo {pages} {1111} (\bibinfo {year} {2007})}\BibitemShut {NoStop}%
\bibitem [{\citenamefont {Georges}\ \emph {et~al.}(1996)\citenamefont
  {Georges}, \citenamefont {Kotliar}, \citenamefont {Krauth},\ and\
  \citenamefont {Rozenberg}}]{RevModPhys.68.13}%
  \BibitemOpen
  \bibfield  {author} {\bibinfo {author} {\bibfnamefont {A.}~\bibnamefont
  {Georges}}, \bibinfo {author} {\bibfnamefont {G.}~\bibnamefont {Kotliar}},
  \bibinfo {author} {\bibfnamefont {W.}~\bibnamefont {Krauth}}, \ and\ \bibinfo
  {author} {\bibfnamefont {M.~J.}\ \bibnamefont {Rozenberg}},\ }\href {\doibase
  10.1103/RevModPhys.68.13} {\bibfield  {journal} {\bibinfo  {journal} {Rev.
  Mod. Phys.}\ }\textbf {\bibinfo {volume} {68}},\ \bibinfo {pages} {13}
  (\bibinfo {year} {1996})}\BibitemShut {NoStop}%
\bibitem [{\citenamefont {Bauer}\ \emph {et~al.}(2011)\citenamefont {Bauer},
  \citenamefont {Carr}, \citenamefont {Evertz}, \citenamefont {Feiguin},
  \citenamefont {Freire}, \citenamefont {Fuchs}, \citenamefont {Gamper},
  \citenamefont {Gukelberger}, \citenamefont {Gull}, \citenamefont {Guertler},
  \citenamefont {Hehn}, \citenamefont {Igarashi}, \citenamefont {Isakov},
  \citenamefont {Koop}, \citenamefont {Ma}, \citenamefont {Mates},
  \citenamefont {Matsuo}, \citenamefont {Parcollet}, \citenamefont {Pawlowski},
  \citenamefont {Picon}, \citenamefont {Pollet}, \citenamefont {Santos},
  \citenamefont {Scarola}, \citenamefont {Schollwock}, \citenamefont {Silva},
  \citenamefont {Surer}, \citenamefont {Todo}, \citenamefont {Trebst},
  \citenamefont {Troyer}, \citenamefont {Wall}, \citenamefont {Werner},\ and\
  \citenamefont {Wessel}}]{BBauer:2011tz}%
  \BibitemOpen
  \bibfield  {author} {\bibinfo {author} {\bibfnamefont {B.}~\bibnamefont
  {Bauer}}, \bibinfo {author} {\bibfnamefont {L.~D.}\ \bibnamefont {Carr}},
  \bibinfo {author} {\bibfnamefont {H.~G.}\ \bibnamefont {Evertz}}, \bibinfo
  {author} {\bibfnamefont {A.}~\bibnamefont {Feiguin}}, \bibinfo {author}
  {\bibfnamefont {J.}~\bibnamefont {Freire}}, \bibinfo {author} {\bibfnamefont
  {S.}~\bibnamefont {Fuchs}}, \bibinfo {author} {\bibfnamefont
  {L.}~\bibnamefont {Gamper}}, \bibinfo {author} {\bibfnamefont
  {J.}~\bibnamefont {Gukelberger}}, \bibinfo {author} {\bibfnamefont
  {E.}~\bibnamefont {Gull}}, \bibinfo {author} {\bibfnamefont {S.}~\bibnamefont
  {Guertler}}, \bibinfo {author} {\bibfnamefont {A.}~\bibnamefont {Hehn}},
  \bibinfo {author} {\bibfnamefont {R.}~\bibnamefont {Igarashi}}, \bibinfo
  {author} {\bibfnamefont {S.~V.}\ \bibnamefont {Isakov}}, \bibinfo {author}
  {\bibfnamefont {D.}~\bibnamefont {Koop}}, \bibinfo {author} {\bibfnamefont
  {P.~N.}\ \bibnamefont {Ma}}, \bibinfo {author} {\bibfnamefont
  {P.}~\bibnamefont {Mates}}, \bibinfo {author} {\bibfnamefont
  {H.}~\bibnamefont {Matsuo}}, \bibinfo {author} {\bibfnamefont
  {O.}~\bibnamefont {Parcollet}}, \bibinfo {author} {\bibfnamefont
  {G.}~\bibnamefont {Pawlowski}}, \bibinfo {author} {\bibfnamefont {J.~D.}\
  \bibnamefont {Picon}}, \bibinfo {author} {\bibfnamefont {L.}~\bibnamefont
  {Pollet}}, \bibinfo {author} {\bibfnamefont {E.}~\bibnamefont {Santos}},
  \bibinfo {author} {\bibfnamefont {V.~W.}\ \bibnamefont {Scarola}}, \bibinfo
  {author} {\bibfnamefont {U.}~\bibnamefont {Schollwock}}, \bibinfo {author}
  {\bibfnamefont {C.}~\bibnamefont {Silva}}, \bibinfo {author} {\bibfnamefont
  {B.}~\bibnamefont {Surer}}, \bibinfo {author} {\bibfnamefont
  {S.}~\bibnamefont {Todo}}, \bibinfo {author} {\bibfnamefont {S.}~\bibnamefont
  {Trebst}}, \bibinfo {author} {\bibfnamefont {M.}~\bibnamefont {Troyer}},
  \bibinfo {author} {\bibfnamefont {M.~L.}\ \bibnamefont {Wall}}, \bibinfo
  {author} {\bibfnamefont {P.}~\bibnamefont {Werner}}, \ and\ \bibinfo {author}
  {\bibfnamefont {S.}~\bibnamefont {Wessel}},\ }\href
  {http://iopscience.iop.org/1742-5468/2011/05/P05001} {\bibfield  {journal}
  {\bibinfo  {journal} {J. Stat. Mech.: Theor. Exp.}\ }\textbf {\bibinfo
  {volume} {2011}},\ \bibinfo {pages} {P05001} (\bibinfo {year}
  {2011})}\BibitemShut {NoStop}%
\bibitem [{\citenamefont {Assaad}\ and\ \citenamefont
  {Evertz}(2008)}]{Assaad:2008hx}%
  \BibitemOpen
  \bibfield  {author} {\bibinfo {author} {\bibfnamefont {F.~F.}\ \bibnamefont
  {Assaad}}\ and\ \bibinfo {author} {\bibfnamefont {H.~G.}\ \bibnamefont
  {Evertz}},\ }\href
  {http://link.springer.com/chapter/10.1007/978-3-540-74686-7_10} {\bibfield
  {journal} {\bibinfo  {journal} {Computational Many Particle Physics, Lecture
  Notes in Physics}\ }\textbf {\bibinfo {volume} {739}},\ \bibinfo {pages}
  {277} (\bibinfo {year} {2008})}\BibitemShut {NoStop}%
\bibitem [{\citenamefont {Hirsch}(1988)}]{Hirsch:1988wo}%
  \BibitemOpen
  \bibfield  {author} {\bibinfo {author} {\bibfnamefont {J.}~\bibnamefont
  {Hirsch}},\ }\href
  {http://eutils.ncbi.nlm.nih.gov/entrez/eutils/elink.fcgi?dbfrom=pubmed&id=9946131&retmode=ref&cmd=prlinks}
  {\bibfield  {journal} {\bibinfo  {journal} {Phys. Rev., B Condens. Matter}\
  }\textbf {\bibinfo {volume} {38}},\ \bibinfo {pages} {12023} (\bibinfo {year}
  {1988})}\BibitemShut {NoStop}%
\end{thebibliography}%


\clearpage 

\appendix
\section{Supplementary Materials}

\subsection{Renyi Entanglement Entropy of Noninteracting Fermions}

The reduced density matrix $\hat{\rho}_{A}$ of a free fermion system is fully determined by the correlation matrix~\cite{Peschel:2002gz}

\begin{equation*}
C_{\mathbf{ij}} = \left\langle  \hat{c}^{\dagger}_\mathbf{i} \hat{c}_\mathbf{j} \right\rangle,
\end{equation*}
where $\langle \ldots \rangle $ denotes thermal average with respect the noninteracting Hamiltonian. To calculate the Renyi entanglement entropy, we restrict $C_{\mathbf{ij}}$ to a subregion A and diagonalize it to get eigenvalues $\zeta_{\ell}$. Introducing $\xi_{\ell} = \ln(\frac{1-\zeta_{\ell}}{\zeta_{\ell}})$, the rank-$n$ Renyi entanglement entropy of free fermions is calculated as 

\begin{equation*}
S_{n}^{0} = \frac{1}{1-n} \sum_{\ell} \left[ \ln (1+e^{-n\xi_{\ell}}) - n \ln(1+e^{-\xi_{\ell}}) \right]. 
\end{equation*}

\subsection{Noninteracting Green's Functions}
The Green's function matrices appeared in Eqs.~(\ref{eq:ZAweight}-\ref{eq:Z2weight}) reads
\begin{eqnarray}
\left(G^{k}_{\mathbb{X}} \right) _{pq} & =&\mathcal{G}^{0}_{\mathbb{X}}(\tau_{p}, \tau_{q}) _{\mathbf{i}_{p}\mathbf{i}_{q} } - \delta_{pq}/2,
\label{eq:G}
\end{eqnarray}
where $\mathbb{X}\in \{ \mathcal{Z}^{A}, \mathcal{Z}\}$,  $1<p,q<2k$ are the vertex indices and $\mathbf{i}_{p},\mathbf{i}_{q}$ are the site indices. The term $\delta_{pq}/2$ arises from the constant shift defined in the interaction term. For $\mathbb{X}=\mathcal{Z}^{A}$ and $\tau_{p(q)} \ge \beta$, the corresponding site index $\mathbf{i}_{p(q)}$ is interpreted  as a site in the artificially enlarged part of the system~\cite{Broecker:2014ud}. $\mathcal{G}^{0}_{\mathcal{Z}^{A}}$ and $\mathcal{G}^{0}_{\mathcal{Z}}$ are the noninteracting Green's functions for the $\mathcal{Z}^{A}$  and $\mathcal{Z}$ ensembles. They have the following general form
\cite{Assaad:2008hx}

\begin{widetext}
\begin{equation}
\mathcal{G}^{0}_{\mathbb{X}}(\tau_{1}, \tau_{2})   =\begin{cases}
 \mathcal{B}(\tau_{1}, \tau_{2}) [\mathbb{I} + \mathcal{B}(\tau_{2}, 0)\mathcal{B}(\tau_{\max}, \tau_{2}) ]^{-1} , & \text{$\tau_{1}\ge \tau_{2}$}, \\ \\
 \{ [\mathbb{I} +    \mathcal{B}(\tau_{1}, 0)\mathcal{B}(\tau_{\max}, \tau_{1})]^{-1} -\mathbb{I}\}   \mathcal{B}^{-1}(\tau_{2},\tau_{1})  
    , & \text{$\tau_{1}< \tau_{2}$}.
  \end{cases}
\label{eq:nonintgf}  
\end{equation}
\end{widetext}
where $\mathcal{B}(\tau_{1}, \tau_{2})= \mathcal{T} [e^{-\int_{\tau_{2}}^{\tau_{1}} \mathrm{d}\tau K(\tau)}]$ is the noninteracting time-ordered propagator, with the matrix $K$ being the quadratic part $\sum_\mathbf{i,j} \hat{c}^{\dagger}_{\mathbf{i}} K_\mathbf{ij} \hat{c}_{\mathbf{j}}$ of the Hamiltonian. 
 
For $\mathbb{X} = \mathcal{Z}$ one has $\tau_{\max}=\beta$ and the fact that $K$ is imaginary-time independent allows simplification of Eq.~({\ref{eq:nonintgf}}),  

\begin{equation}
\mathcal{G}^{0}_{\mathcal{Z}}(\tau_{1},\tau_{2})  =\begin{cases}
 \frac{e^{-K(\tau_{1}-\tau_{2})}}{1+e^{- \beta K}}, & \text{$\tau_{1}\ge \tau_{2}$}, \\ \\
    -\frac{e^{-K(\tau_{1}-\tau_{2})}}{1+e^{\beta K}}, & \text{$\tau_{1}< \tau_{2}$}.
  \end{cases}
\end{equation}
Since $\mathcal{G}^{0}_{\mathcal{Z}}(\tau_{1},\tau_{2})$  only depends the imaginary-time difference $\tau_{1}-\tau_{2}$ we precompute it on a fine $\tau$-mesh and use linear interpolations to get the value of $\mathcal{G}^{0}_{\mathcal{Z}}(\tau_{1}, \tau_{2})$ for arbitrary imaginary-times.

For $\mathbb{X} = \mathcal{Z}^{A}$ one has $\tau_{\max}=2\beta$ and $K(\tau)$ is an artificial imaginary-time dependent Hamiltonian~\cite{Broecker:2014ud}. In general $\mathcal{G}^{0}_{\mathcal{Z}^{A}}(\tau_{1}, \tau_{2})$ depends both on $\tau_{1}$ and $\tau_{2}$. We can also precompute $\mathcal{G}^{0}_{\mathcal{Z}^{A}}(\tau_{1}, \tau_{2})$ on a fine mesh and perform bilinear interpolations. However, storing $\mathcal{G}^{0}_{\mathcal{Z}^{A}}(\tau_{1}, \tau_{2})$  is not possible for large system size, and as a compromise we only pre-calculate $\mathcal{G}^{0}_{ \mathcal{Z}^{A}}$ on a 
coarse imaginary-time grid $(\bar{\tau}_{1}, \bar{\tau}_{2}) $ using Hirsch's matrix inversion method~\cite{Hirsch:1988wo}. In the CTQMC calculation, whenever a new vertex is proposed, we find the nearest imaginary time point  $(\bar{\tau}_{1},\bar{\tau}_{2})$ such that $\tau_{1}\ge \bar{\tau}_{1}$ and $\tau_{2}\ge \bar{\tau}_{2}$  and compute the required Green's function using 

\begin{equation}
\mathcal{G}^{0}_{ \mathcal{Z}^{A}}(\tau_{1}, \tau_{2})  =  \mathcal{B}(\tau_{1}, \bar{\tau}_{1}) \mathcal{G}^{0}_{ \mathcal{Z}^{A}}(\bar{\tau}_{1}, \bar{\tau}_{2}) \mathcal{B}^{-1}(\tau_{2}, \bar{\tau}_{2}) 
\end{equation}
In this way we avoid CPU expensive matrix inversion in Eq.~(\ref{eq:nonintgf}). 

\end{document}